\documentclass[twocolumn]{revtex4}
\usepackage{latexsym}
\usepackage{amsmath}
\usepackage{times}
\usepackage{amssymb}
\usepackage{fancyheadings}
\usepackage[T1]{fontenc}
\usepackage[utf8]{inputenc}
\usepackage{url}
\usepackage{hyperref}
\usepackage{color}
\usepackage{graphicx}
 \usepackage{epsfig}
\usepackage{mathptmx}
\usepackage{bm}
\usepackage{array}
\usepackage{algorithm}
\usepackage{algorithmic}
\usepackage{url}
\usepackage{hyperref}
\usepackage{algorithm}
\usepackage{algorithmic}
\usepackage{color}

\usepackage{lscape}

\usepackage{draftcopy}

\begin{document}
\title{Multifractal analysis of Barkhausen noise reveals the dynamic nature of criticality at hysteresis loop}
\author{Bosiljka Tadi\'c}
\affiliation{ 
Department  for Theoretical Physics; Jo\v{z}ef Stefan Institute; 
P.O. Box 3000; SI-1001 Ljubljana; Slovenia\\ \hspace{1cm}} 

\begin{abstract} The field-driven magnetisation reversal processes in disordered systems exhibit a collective behaviour that is manifested in the scale-invariance of avalanches,  closely related to underlying dynamical mechanisms. Using the multifractal time series analysis, we study the structure of fluctuations at different scales in the accompanying   Barkhausen noise. The stochastic signal represents the magnetisation discontinuities along the hysteresis loop of a  3-dimensional random field Ising model simulated for varied disorder strength and driving rates. The analysis of the spectrum of the generalised Hurst exponents reveals that the segments of the signal with large fluctuations represent two distinct classes of stochastic processes in weak and strong pinning regimes.   Furthermore, increased driving rates have a profound effect on the small fluctuation segments and broadening of the spectrum.  The study of the temporal correlations, sequences of avalanches, and their scaling features complements the quantitative measures of the collective dynamics at the hysteresis loop. The multifractal properties of Barkhausen noise describe the dynamical state of domains and precisely discriminate the weak pinning,  permitting the motion of individual walls, from the mechanisms occurring in strongly disordered systems. The multifractal nature of the reversal processes is particularly relevant for currently investigated memory devices that utilize  a controlled motion of individual domain walls.
\end{abstract}

\maketitle

 \section{Introduction}
In driven disordered spin systems, the form of hysteresis loop reflects the existence of the domain structure, which responds to the external magnetic field by a motion of the domain walls. In particular, an increase of the domains aligned with the field occurs while the opposite magnetisation domains shrink. Thus, the magnetisation reversal processes in these systems involve a complex interplay between the motion of the domain walls and their pinning by the magnetic and structural disorder centers.  Driven by a slow field ramping along the hysteresis loop,  the magnetisation reversal exhibits avalanches of aligned spins when the domain wall moves to a new position. The resulting burst events of magnetisation jumps is known as Barkhausen noise. These avalanches exhibit scaling features that depend on the strength of pinning and the driving rate, closely reflecting the underlying dynamical mechanism. 
In this work, we use the multifractal analysis of the Barkhausen noise to investigate the nature of fluctuations at all scales, which characterise the active mechanisms in different pinning regimes.

Ferromagnetic alloys and metallic glasses exhibiting Barkhausen effect \cite{Exp_review_RModPhys1953,durin_expBHN1996,djole_expBHN1996,alloys2003} represent strongly disordered systems with domain structure grafted by the fabrication. Similar phenomena  occur in  relaxor ferroelectrics \cite{relaxorferroelectrics_BHN2002}, stress-induced martensites  \cite{martensites}, porous media  \cite{eduard_porous2014} and other systems with a hysteresis.
These phenomena represent a fascinating physics problem. Besides,  the occurrence of hysteresis loop and the magnetisation reversal processes provide the basis for noninvasive structural analysis of materials \cite{russJnondistrTesting_reviewBHN2000} and technological applications. Prominent examples are the magnetic memory and, connected with charge transport, spin electronics in data storage \cite{Nat_spintronics}.
Recently proposed memory devices are based on the controlled manipulation of the domain-wall motion and pinning by weak disorder or geometrical constraints in magnetic nanowires \cite{science-rew2008}. A direct observation of the domain-wall motion in different experimental settings  \cite{AFM_BHN2004,DWstochasticity2011,DWstochasticity_NatComm2010,DWstochasticity_expresistanceNoise2010} revealed their stochastic kinetics, exhibiting certain universal features and dependence on the type and density of pinning centers. Consequently, a deeper understanding of the stochastic processes on the hysteresis loop becomes a topic of increased interest for these systems.

Robust scale invariance of the magnetisation reversal avalanches in disordered ferromagnets has been observed both in experiments and numerical simulations \cite{durin_fractalBHN1995,berger_expBHN2000,durin_expPRL2000,cornell_perkovic1999,eduard_anisotropy2001}.  
The scaling behaviour of  Barkhausen avalanches resembles  familiar critical phenomena \cite{cornell_perkovic1999}, while the dynamics of domain walls  and spin flips can be related with criticality in sandpile automata \cite{sotolongo_SPAbhn2000} or bootstrap percolation \cite{DD_bootstrappercol2002}. Moreover, it shares some general features with the collective dynamical phenomena  observed at fixed points by the  renormalization-group analysis of model disordered systems \cite{RenormGroup-comb}. 
The scale invariance  of the reversal avalanches depends on the system's dimensionality, range of interactions and the spin symmetry.  In addition, these dynamical phenomena on the hysteresis were shown to depend on the magnetic anisotropy and strength of disorder \cite{eduard_anisotropy2001,BT_PRL1996},  topology of the substrate \cite{DD_Bethelatt1-weNet}, as well as the  driving mode \cite{we_drivingrate2004,bahiana_drivingrates2001}.
The occurrence of a variety of the exponents (a summary of the exponents can be found in Refs.  \cite{BT_PhysA1999,eduard_anisotropy2001}) and universality classes in the scaling of avalanches is believed to strictly reflect differences in the underlying dynamical mechanisms \cite{BCN_2015}. 
 Specifically, for the weak disorder the system spanning avalanches can occur, whose fronts represent moving individual domain walls. On the other hand, the occurrence of many domains at strong disorder makes the mutually constrained motion of many domain walls that results in small (subcritical) avalanches. The transition between these regimes has been considered as a disorder-induced critical  point \cite{cornell_perkovic1999,bhn-critdisorder2x}.
While the scaling of the spanning avalanches has been studied numerically with an astonishing precision \cite{eduard_FSS2003,eduard_spanning2004,djole_scaling2D,djole_spanningaval2014}, there is little knowledge about the fine-scale structure of the collective fluctuations, which are encoded in the accompanying Barkhausen noise. 

Here, we study the multifractal properties of Barkhausen noise signals across a broad range of pinning and driving conditions. We simulate the processes of magnetisation reversal along the hysteresis loop using  3-dimensional random-field Ising model for the varied strength of disorder and different rates of the field ramping. The analysis reveals that the Barkhausen noise exhibits a multifractal structure, which can be appropriately described by a spectrum of the generalised Hurst exponents. Practically, this means that the fluctuations of the magnetisation during the reversal process decompose into fractal components, each of which exhibits a different scale invariance.  As a complex signal, the Barkhausen noise also shows the temporal correlations, avalanching, and non-Gaussian relaxation or avalanche returns.  It appears that the multifractal spectrum represents a more sensitive measure of the effects of disorder strengths and changes in the driving conditions than the standardly analysed scaling of avalanches. The studied multifractal properties of the Barkhausen noise provide  indicators of the dynamical state of the domain walls in memory materials under experimentally relevant regimes, which combine pinning and driving.

The rest of the paper is organised as follows. In Section\ \ref{sec-TS} we describe the model and simulations and present the evidence that the collective fluctuations are occurring at the hysteresis loop and change with the strength of disorder and driving rate.  The detailed detrended multifractal analysis of the signals for different disorder and driving rates is given in Section\ \ref{sec-MFRA}. The complementary analysis of the scaling of avalanches and temporal correlations of the studied signals is provided in Section\ \ref{sec-avalanches}. Section\ \ref{sec-conclusions} contains a brief summary and a discussion of the results.

\section{Collective fluctuations in the magnetization reversal processes\label{sec-TS}}
We consider the magnetisation reversal under a slow ramping of the field along the hysteresis loop of the random-field Ising model with discrete spin states $S_i=\pm 1$. It is usually assumed in theoretical considerations and numerical simulations that  the effects of magnetic disorder are suitably captured by quenched Gaussian random fields \cite{bhn-critdisorder2x,eduard_FSS2003,eduard_spanning2004,djole_scaling2D,djole_spanningaval2014}.
Recently, a possible realisation of the magnetic system with a tunable random-field disorder has been suggested: a variable transverse field is applied to magnets with a strong spin anisotropy, and driven by a parallel field \cite{RFrealization_Nat2007}. 
Within this context, the random-field Ising model suitably captures the interplay between the magnetic disorder, represented by the quenched random field at each lattice site,  spin--spin interactions, and the driving external magnetic field.  It suffices to consider the zero temperature dynamics, where the spin aligns along the local field $h_i^{loc}(t)$ at time $t$ to minimize the energy, i.e., 
\begin{equation}
S_i(t+1) = sign\left( h_i^{loc}(t)\right) .
\label{eq-si}
\end{equation}
Apart from the quenched random field $h_i$,  the value of the local field at site $i$ consists of the  dynamical contributions given by the current states of the neighbouring spins  and the value of the external field, $B(t)$, which acts at all sites,
\begin{equation} 
h_i^{loc}(t) =\sum_{j\in nn}J_{ij}S_j(t) +h_i +B(t) .
\label{eq-hloc}
\end{equation}
Here, we consider strictly ferromagnetic interactions $J_{ij}=J_0$ for pairs $\{ij\}$ of neighbouring spins on a three-dimensional cubic lattice. The linear lattice size is $L=50$ and at each site $i=1,2,\cdots N =125000$ a quenched random field  $h_i$ is taken from Gaussian distribution $h_i\in P(h,f)$ with a zero mean and the variance $f$ (in units of $J_0^2$).  The presence of site defects \cite{BT_PRL1996} is also considered when $c>0$, where $c$  is the randomly selected fraction of sites without a spin.

\begin{figure}
\resizebox{18pc}{!}{\includegraphics{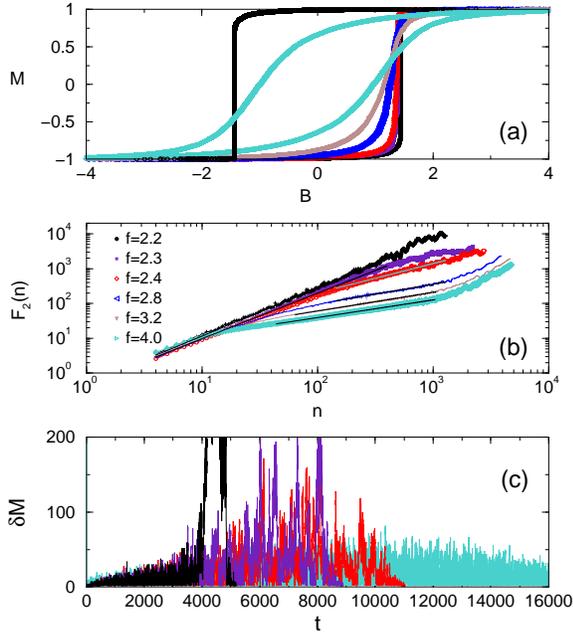}}
\caption{\small (a) Magnetisation $M$ (normalised by the saturation value $M_{sat}=\sum_iS_i$) plotted against the external field $B$ for different disorder $f$ indicated in the legend. The shown complete loops corresponds  the weakest (square loop) and the strongest disorder (S-shaped loop). (b) The standard deviation $F_2(n)$ of the fluctuations at the segment of length $n$, plotted against the segment length for the indicated values of disorder, corresponding to the hysteresis loops in top panel. 
(c) Examples of Barkhausen jumps $\delta M$ against time steps $t$ for strong disorder $f=4.0$ (background line) and  weak disorder $f=$2.4, 2.3 and 2.2, corresponding to gradually shorter signals.}
\label{fig-BHN-hl}
\end{figure}

The periodic boundary conditions are applied in all directions, and the parallel update of all spins is performed at each time step. That is, at each step $t$ the local fields are computed at all sites and spins updated by aligning each spin according to the sign of the corresponding local field. The number of the flipped spins at a particular time step $t$ comprises the magnetisation discontinuity $\delta M$ at time $t$, or the data point of the Barkhausen noise. The process starts from the homogeneous state with all spins down and a large negative external field $B(0) =-B_{max}$. Then the field is adiabatically increased over time, as described below, until the positive value $B_{max}$ is reached.
The field is increased by a small amount $r\equiv \Delta B/J_0$ and kept constant until the system completely relaxes in the current field. Thus, during one field rump, the system can experience a cascade of spin alignments, which can last for several time steps. All spins aligned parallel to the field during one field ramp define the size $s$ of an avalanche while the number of steps comprises the avalanche duration $T$.
In the corresponding Barkhausen noise signal, Fig.\ \ref{fig-BHN-hl}c, an avalanche comprises the magnetisation bursts between two consecutive drops of the signal to the zero level. After the system has relaxed, the field is increased again.
In this way, the field increases according to $B(t+T_t) =B(t) + \Delta B$, where $T_t$ is the duration of the avalanche triggered at the moment $t$. Therefore, the effective rate  $r_{eff}=\langle dB(t)/dt\rangle$ can be much lower than the applied driving rate $r$  in the middle of the loop where the large avalanches may occur.

\begin{figure}[htb]
\begin{tabular}{cc} 
\resizebox{18pc}{!}{\includegraphics{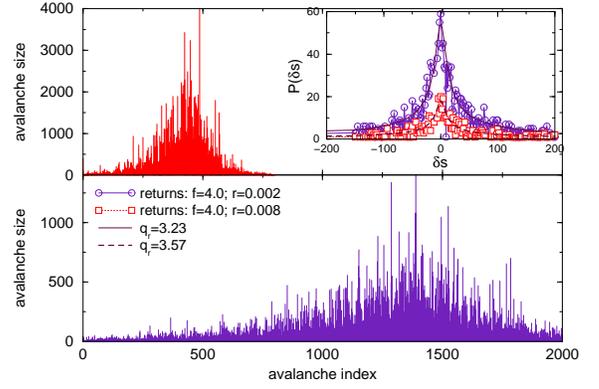}}\\
\end{tabular}
\caption{\small Sequence of avalanches occurring in the ascending branch of the hysteresis for strong disorder $f=$4.0 and slow driving  $r=0.002$ (lower panel) and fast driving $r= 0.008$ (top panel). Inset: The corresponding distributions of the avalanche returns. Fits with the expression (\ref{eq-gGauss}): $A=47.9\pm 2.2$, $D=13.56\pm1.71$, $q_r=3.23\pm 0.02$, for the low driving rate $r=0.002$, and  $A=16.86\pm 0.98$, $D=10.1\pm1.2$, $q_r=3.57\pm 0.21$, for $r=0.008$.}
\label{fig-returns-PS}
\end{figure}

In the simulations, we set  $J_0=1$ and $c=0$ unless otherwise stated; also,  $B_{max}=-6$  or, otherwise,  the value that matches the minimum value of all random fields in the current sample. We record the noise signal from the complete half-loop in the first sample. The cumulative distributions of the avalanches are computed and averaged over ten samples with a different allocation of random fields.   Fig.\ \ref{fig-BHN-hl} illustrates the effects of disorder on the hysteresis loop and the properties of the signal. 
The examples of the Barkhausen noise signal shown in Fig.\ \ref{fig-BHN-hl}c corresponding to the ascending branches of the hysteresis loop for several values of the disorder $f$ and a low driving rate $r=0.002$. 
The all considered strengths of disorder measured by the parameter $f$ were indicated in the legend of Fig.\ \ref{fig-BHN-hl}. By increased disorder, the corresponding hysteresis loop shrinks, top panel. At the same time, the scaling features of the standard fluctuations function $F_2(n)$ of the noise signal around the local trend changes. In particular, the scaling exponent $H_2$, given by the slope indicated on the curves $F_2(n)$  plotted against the segment $n$ in Fig.\ \ref{fig-BHN-hl}b, decreases from  $H_2 $= 1.347, 1.326 for weak disorder $f=$2.2, 2.3, via $H_2=$0.844  for $f=$2.4, to  $H_2 =$0.571, 0.537 and 0.498  within numerical error bars $\pm0.06$, in the strong disorder $f=$2.8, 3.2, and 4.0, respectively. In section \ref{sec-MFRA}, we will consider the entire spectrum of the generalised Hurst exponents as a marked signature of the signal's complexity, related to the cooperative pinning and driving. Other profound features of the collective dynamics are the occurrence of temporal correlations of the signal itself as well as the clustering of events (avalanches), studied in section\ \ref{sec-avalanches}. In the remaining part of this section, we analyse the sequence of avalanches, which provides the system's response at different driving rates.  The results for a fixed disorder $f=4.0$ are shown in Fig.\ \ref{fig-returns-PS}.  

In complex systems, the sequence of avalanches as they occur in time contains a relevant information about the degree of complexity of the underlying relaxation processes \cite{pavlos_q-returns2014}.  The \textit{avalanche return} is defined as the difference between sizes of two consecutive avalanches in the temporal sequences, 
\begin{equation}
\delta s_\lambda = s_{\lambda +1} - s_\lambda 
\label{eq-return}
\end{equation} 
where  $\lambda$ is the avalanche index in the considered sequence and $s_\lambda$ represents the size of the avalanche.
 As Fig.\ \ref{fig-returns-PS} shows, a different sequence of avalanches is found for the same disorder but increased driving rate. For the strong disorder, the small avalanches occur, corresponding to the limited motion of many domain walls. Consequently, the sequence of avalanches that complete the ascending branch of the hysteresis loop is rather long. By the increased driving rate, however, larger avalanches can occur due to overdriving of weaker pinning centers. Consequently, the number of avalanches corresponding to the complete reversal is smaller (Fig.\ \ref{fig-returns-PS} top panel).
The distribution of the avalanche returns, $P(\delta s)$,  shows wings for large positive/negative deviations, indicating non-Gaussian fluctuations in the avalanche sizes. In the present context, this means that a small variation in the driving field can trigger an avalanche of entirely different size. Such situations mostly occur in the middle of the loop, near the coercive field. 
The distributions can be fitted with the $q_r$-Gaussian function
\begin{equation}
P(\delta s) =A\left[1-(1-q_r)\left(\frac{\delta s}{D}\right)^2\right]^{1/1-q_r}
\label{eq-gGauss}
\end{equation}
which is characteristic of the systems with collective dynamical effects  (for a review, see   \cite{pavlos_q-returns2014,tsallis-book2004}  and references there). Here, the relaxation parameter increases from $q_r=3.23\pm 0.02$, for the case of low driving rate, to  $q_r=3.57\pm 0.21$, for the large driving rate, indicating different propagation of the avalanches.

\section{Detrended multifractal analysis of Barkhausen time series\label{sec-MFRA}}
In analogy to geometrical fractals, time series representing fluctuations of an observable in many complex systems may exhibit power-law singularities \cite{MFRA-uspekhi2007,DMFRA2002}. In particular,  in the vicinity of a point $t$, the variation of the data $\vert \nabla n(t,\epsilon)\vert _{\epsilon \to 0}\sim \epsilon ^{\alpha (t)}$ with an exponent depending on the data point \cite{MFRA-uspekhi2007,DMFRA2002}.  A smaller value of $\alpha (t)$ indicates a stronger irregularity of the signal at that point.
Consequently, these complex signals are described by a singularity spectrum $\psi (\alpha)$ representing a fractal dimension of the subset of the time series with the same singularity exponent $\alpha$. From the point of view of self-similarity, such complex signals are only locally self-similar. 
Practically, this means that different segments of the time series need to be amplified in a particular way to becoming similar to the whole signal. The singularity spectrum is thus used to characterise the nature of the stochastic process and, consequently, to classify the states of the dynamical system. A classical example in physics is the signal of velocity fluctuations in fully developed turbulence \cite{mfr-turbulence2013}. Recently, quantum multifractality reflecting  different scales in the wave function near disorder-induced metal--insulator transition has been studied \cite{qmfr-Andersontransition2015}. Multifractality has been used to characterise  complex signals coming from many real-world systems, both natural and social  \cite{MFRA-uspekhi2007,mfr-review2009,dmfra-sunspot2006,mfr-magnetorheology2014}.

There are several mathematically equivalent ways to determine the singularity spectrum. For practical purposes, the method described in \cite{MFRA-uspekhi2007,DMFRA2002} that utilizes the underlying self-similarity is suitable for the analysis of Barkhausen signals and will be used here. In this approach, the generalised Hurst exponent $H(q)$ as a function of the amplification parameter $q$ is determined numerically, as described below. To suitably amplify different sections of the analysed time series, the parameter $q\in \cal{R}$ takes a range of real values.  It is shown \cite{MFRA-uspekhi2007,DMFRA2002} that  the scaling relation $\tau (q)=qH(q)-1$ holds, where $\tau (q)$ is the exponent of the standard measure (box probability)  defined in the partition function method.  Hence, the spectrum of the generalised Hurst exponents $H(q)$ can be related with the singularity spectrum via Legendre transform  of $\tau (q)$, i.e.,  $\Psi(\alpha)=q\alpha -\tau (q)$, where $\alpha =d\tau/dq = H(q)+qdH/dq$. Obviously, for a monofractal $H(q)=H=const$ and $\alpha = H$, reducing the spectrum $\Psi (\alpha)$ to a single point. 
The functional dependence $H(q)$, where 
\begin{equation}
\delta H=H_{max}(q)-H_{min}(q) \neq 0
\label{eq-widthHq}
\end{equation}
defines the width of the spectrum (the degree of multifractality), also induces a nontrivial singularity spectrum $\Psi (\alpha)$.

\begin{figure}[htb]
\begin{tabular}{cc} 
\resizebox{18pc}{!}{\includegraphics{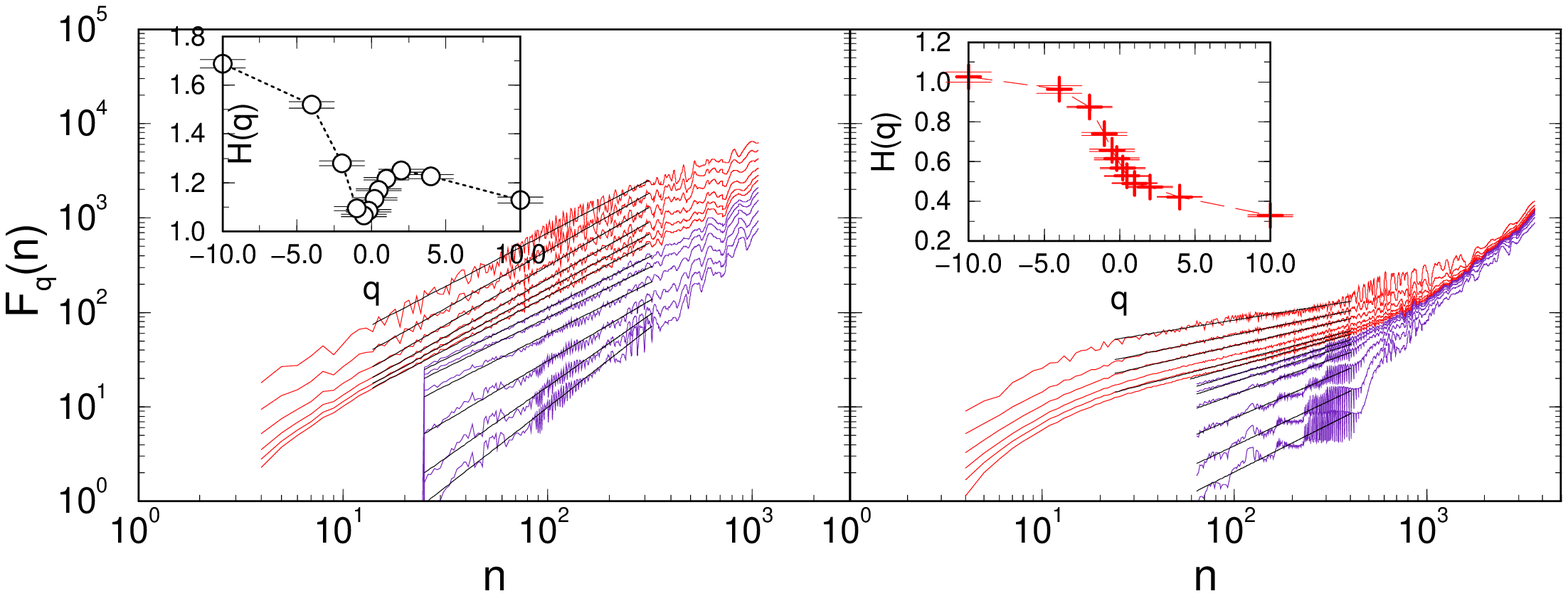}}\\
\resizebox{18pc}{!}{\includegraphics{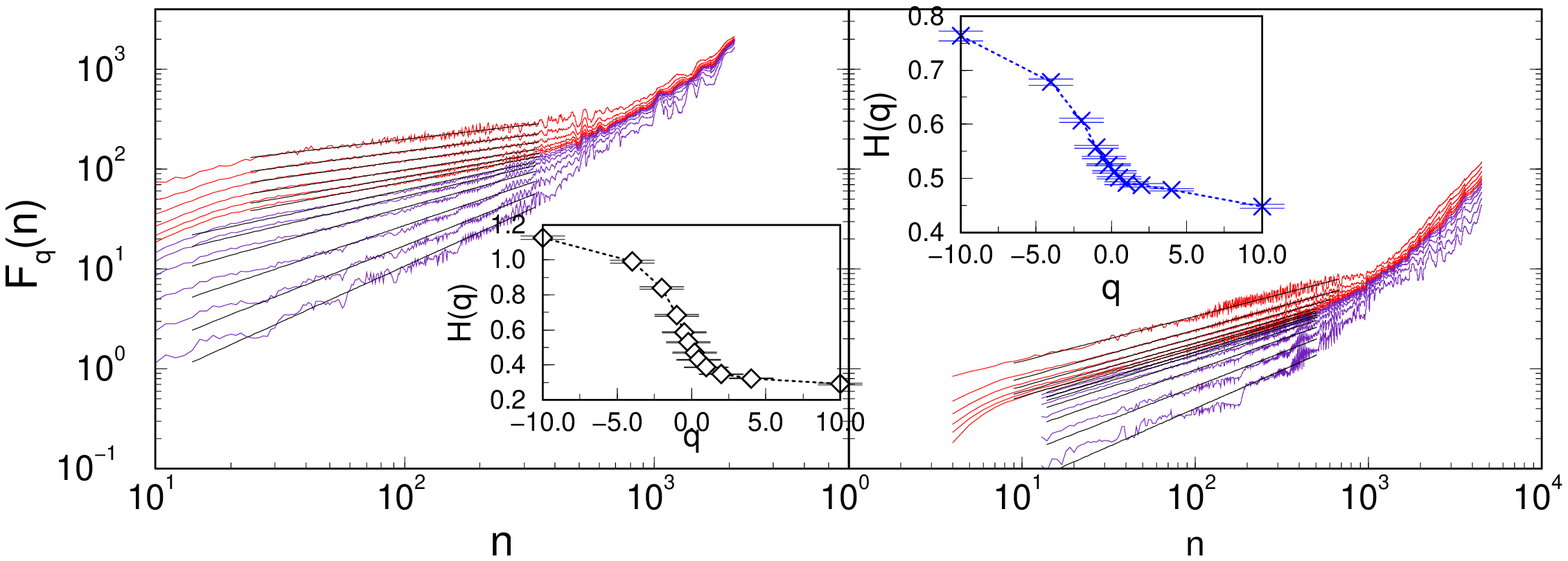}}\\
\end{tabular}
\caption{\small Scaling analysis of the $F_q(n)$ for  Barkhausen time series in weak $f=2.3$ (top panels) and strong  random-field disorder $f=4.0$ (bottom panels); Left panels correspond to the absence of site defects $c=0.0$ and  large driving rate $r=0.008$, while right panels  show the corresponding functions for added site disorder $c=0.3$ and low driving rate $r=0.002$. In each panel, different curves  from bottom to top are for several values of $q\in[-10,+10]$, and the corresponding inset summarises their slopes (fitted range on each curve is indicated by a straight line).}
\label{fig-mfra-fr}
\end{figure}

\begin{figure}[htb]
\begin{tabular}{cc} 
\resizebox{18pc}{!}{\includegraphics{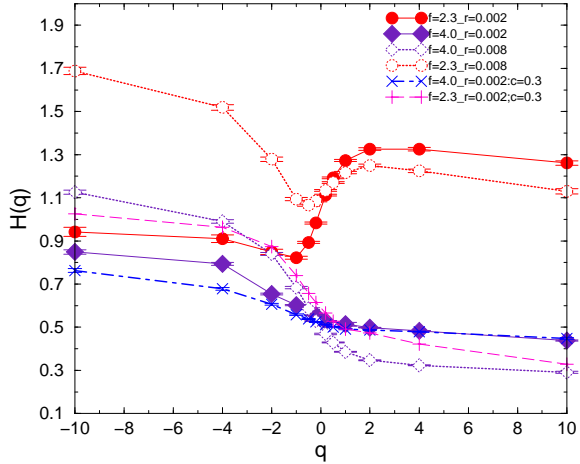}}\\
\end{tabular}
\caption{\small Generalised Hurst exponent $H(q)$ plotted against the distortion parameter $q$  for  random field disorder $f=2.3$  and $f=4.0$  in the absence of site defects $c=0.0$ and two driving rates  $r=0.002$ (solid symbols) $r=0.008$ (empty symbols). The corresponding lines with $+$ and $\times$ symbols are for the addition of site defect, $c=0.3$, in the low driving rate $r=0.002$.}
\label{fig-Hq-fr}
\end{figure}

The time series consists of the magnetisation jumps, $\delta  M(k)$ for time steps $k=1,2,\cdots T_{max}$, where $T_{max}$ is the length of the time series, which  depends on the form of the hysteresis loop. The procedure to determine the generalised Hurst exponent $H(q)$, as described in \cite{MFRA-uspekhi2007,DMFRA2002,dmfra-sunspot2006}, consists of the following steps. 
 First, the profile of the time series is obtained by integration 
\begin{equation}
Y(i) =\sum_{k=1}^i(\delta M(k)-\langle \delta M\rangle)  \ .
\label{eq-profile} 
\end{equation}
The profile is then divided into non-overlapping segments of equal length $n$.  Since the length of the considered time series can vary, the process is repeated starting from the end of the time series, thus in total $2N_s=2Int(T_{max}/n)$ segments are considered.
  Then, at each segment $\mu=1,2\cdots N_s$, the local trend $y_\mu(i)$ is determined (in this case, a linear interpolation is sufficient)  and the standard deviation around  the local trend 
\begin{equation}
 F^2(\mu,n) = \frac{1}{n}\sum_{i=1}^n[Y((\mu-1)n+i)-y_\mu(i)]^2 
\label{eq-F2}
\end{equation}
is determined.  Similarly, $F^2(\mu,n) = \frac{1}{n}\sum_{i=1}^n[Y(N-(\mu-N_s)n+i)-y_\mu(i)]^2$ for $\mu =N_s+1,\cdots 2N_s$. Then, the $q$-th order fluctuation function $F_q(n)$ is obtained  for varied segment length $n$ and averaged over all segments:  
\begin{equation}
F_q(n)=\left\{(1/2N_s)\sum_{\mu=1}^{2N_s} \left[F^2(\mu,n)\right]^{q/2}\right\}^{1/q} \sim n^{H(q)}  \ .
\label{eq-FqHq}
\end{equation}
The scale invariance of the fluctuation function $F_q(n)$ against the segment length $n$ is examined to determine the corresponding scaling exponent $H(q)$. The considered segment lengths vary in the range $n\in[2,int(T_{max}/4)]$.
The distortion parameter $q$ takes a range of real values. In this way, both \textit{small fluctuation segments}, enhanced by the negative values of $q$, and the segments with \textit{large fluctuations}, which dominate the fluctuation function  for the positive values of $q$, are examined. 
In Fig.\ \ref{fig-mfra-fr} the analysis is demonstrated for several examples of the studied  Barkhausen noise series for $q\in[-10,+10]$.  The resulting exponents $H(q)$ for weak and strong pinning strengths and two different driving rates are  plotted in Fig.\ \ref{fig-Hq-fr}. 
Note that the standard fluctuations investigated in Fig.\ \ref{fig-BHN-hl}b for different disorder strengths, result in the exponents $H_2\equiv H(q=2)$, which is directly related to the familiar Hurst exponent $H$. In particular, $0\leq H=H_2\leq 1$ for the fractional Brownian motion, as we found for the profile (\ref{eq-profile}) for $f\geq 2.4$. This implies that the original signal $\delta M(k)$ in the strong disorder represents the fractional Gaussian noise. Whereas $H=H_2-1$ when $H_2>1$, e.g., for the sum of the fractional Brownian processes (see, for instance \cite{dmfra-sunspot2006}), as we find in the case of the weak disorder.

As the Fig.\ \ref{fig-Hq-fr} shows, different $H(q)$ dependences appear to characterise the  multifractality of Barkhausen noise at  weak and strong disorder (cf. the signals in Fig.\ \ref{fig-BHN-hl}c). In particular, for a strong disorder $f=4.0$, a typical decreasing function  $H(q)$ is found. $H(-10)=0.849$ for small-fluctuation segments, reducing to nearly white noise  Hurst exponent $H_2=0.498$ is further decreasing  till $H(+10)=0.411$, for  the large fluctuation segments.  On the other hand, for the weak disorder case, an increasing function $H(q)$ is obtained for the large fluctuation segments reaching the maximum at $H_2=1.326$. 
This type of the spectrum results in combined contributions of different features of the signal from the beginning of the hysteresis loop, which dominates smooth sections, and from the central part of the loop, dominating the strong fluctuations of the single-wall dynamics. 
Hence, the large fluctuation segments of the Barkhausen signal have clearly distinct fractal properties in strong and weak pinning regimes. While the smooth segments of these signals have a similar fractal features by slow driving (cf. $H(-10)=0.849$ for strong, and $H(-10)=0.942$ for the weak disorder, respectively). 

Furthermore, the increased driving rate profoundly affects the small fluctuations, resulting in the increase of $H(q)$ for $q<0$  both for strong and weak pinning. Thus, the smooth sections of the signal become even smoother by the fast driving. The changes are more dramatic in the case of weak pinning, where the values $H(q<0)$ much exceed unity than in the strong pinning. On the other hand, the exponents $H(q>0)$ decrease in the fast driving regime both for strong and weak pinning.  Thus, the rough sections of the signal receive reduced exponents; however, the values remain high $H(q>0)>1$ for the weak pinning.   Interestingly, the fast driving regime leads to the anti-persistence of the large fluctuations in the strong disorder case, $H_2=0.347\pm 0.002$, cf. Fig.\ \ref{fig-Hq-fr}. 
Moreover,  the difference between minimum and maximum value  $\delta H \equiv H_{max}(q)-H_{min}(q)$ (and consequently, the width of the singularity spectrum) increases with the increased driving rate. For example, $\delta H$ increases from 0.504 to 0.622 when the driving rate changes from $r=$0.002 to 0.008 in the weak pinning regime. In the case of strong disorder, the corresponding values are $\delta H=$ 0.411 and  0.835, respectively.

Added site defects have a different effect on domain walls than the magnetic random-field disorder. Here, the absence of a spin at the disorder site weakens the interaction contribution in Eq.\ (\ref{eq-hloc}) to the local fields at nearest neighbour sites. Thus, a defect site acts as a nucleation center for a new domain wall. Consequently, even a weak random field disorder in combination with site defects results in the Barkhausen noise with a 'standard' multifractality, i.e., with a decreasing function $H(q)$. In Fig.\ \ref{fig-Hq-fr}, two curves demonstrate the situation with $c=0.3$ added to the above studied random-field disorder $f=2.3$ and $f=4.0$ in the case of slow driving. The  variations of $H(q)$ from $H(-10)=1.025$ via $H_2=0.528$ to $H(+10)=0.45$ in the weak random field disorder  with site defects resembles the curve for strong random fields without site defects. The changed multifractal spectrum, in this case, can be attributed to the presence of many domain walls, which are nucleated at site defects. In the strong random-filed disorder case, the difference $\delta H$ is further reduced by the presence of site defects. Moreover, the fluctuations corresponding to $q>0$, i.e.,  $H_2=0.501$ and $H(+10)=0.448$ approach the properties of white noise.

\section{Scale invariance of the magnetization bursts and the origin of multifractality\label{sec-avalanches}}
In many complex systems,  the intermittency, power-laws, and long-range temporal correlations are often found together with the multifractality as prominent features of collective fluctuations. As some of these characteristics can also appear in monofractals, researchers have considered their presence or absence as the potential origin of multifractality. Hence, as a dominant cause of the multifractality, some studies give evidence of the ubiquitous occurrence of cascades of events \cite{mfr-review2009}. While others focus on formal reasons for the $1/\nu$-noise correlations \cite{mfr-1of2011} or other stochastic features of the signal \cite{MFRA-uspekhi2007,DMFRA2002}. Recently, an analytical study of fully developed turbulence \cite{mfr-turbulence2013} has revealed that the multifractal properties of the velocity fluctuations can be related to a characteristic mathematical structure of the underlying dynamical equation. 
Here, with the numerical analysis we demonstrate that the studied Barkhausen signals possess the temporal correlations as well as power-law distributions of avalanches and elementary signals. Moreover, their quantitative indicators appear to be different in the case of weak pinning from the ones in the strong pinning regimes. 

\begin{figure}[!htb]
\begin{tabular}{cc} 
\resizebox{18pc}{!}{\includegraphics{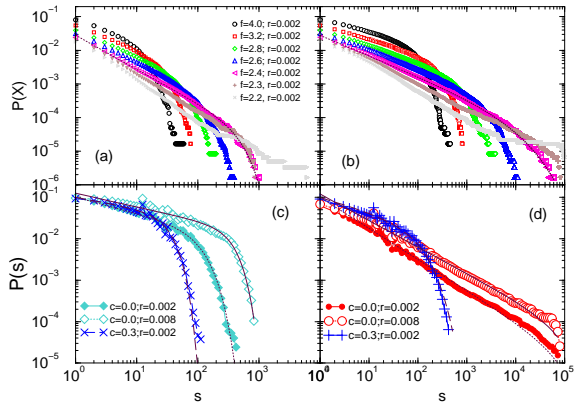}}\\
\end{tabular}
\caption{\small (a) and (b): Cumulative distribution of avalanche durations and sizes, respectively,  for slow driving rate $r$ and different random-field disorder $f$ indicated in the legend.  (c) and (d): Comparative analysis of the avalanche size distributions for $f=4.0$ and   $f=2.3$ , representing strong weak pinning, respectively, for two extreme driving rates $r$. The $\times$ and $+$ symbols correspond to the addition of site defects with concentration $c=0.3$ in the slow driving limit.}
\label{fig-avalanche-distr}
\end{figure}

In Fig.\ \ref{fig-avalanche-distr}, the \textit{cumulative distributions} of the avalanche durations and sizes are determined for different combinations of disorder strengths and driving rates, which  correspond to the Barkhausen nose signals studied by multifractal analysis in Sec.\ \ref{sec-MFRA}.  At low driving rate and the absence of site defects, these distributions strongly depend on the strength of magnetic pinning $f$, in full agreement with  previous studies \cite{cornell_perkovic1999,bhn-critdisorder2x,eduard_FSS2003}. Specifically, system-size avalanches that may occur at weak pinning regime lead to a plateau in the cut-off region. With the increased disorder, a finite cut-off appears, and the cut-off length decreases systematically with the disorder.  The following expression with $\sigma=1$ provides a good fit for the majority of the distributions
\begin{equation}
P(X)=aX^{-b}e^{-\left(\frac{X}{d}\right)^\sigma} \ ,
\label{eq-plexp}
\end{equation}
where $X$ stands for $T$--duration or $s$--size of avalanches. Specifically, the fit lines which are shown in Fig.\ref{fig-avalanche-distr}(a) and (b) correspond to the disorder strength $f=2.4$. The fit parameters are,  $a=0.025\pm 0.002$, $b=0.962\pm 0.002$, $d=336\pm 4$, for $P(T)$, and  $a=0.037\pm 0.002$, $b=0.695\pm 0.001$, $d=19209\pm 39$, for $P(s)$, respectively. 
Similarly, the distributions of avalanche sizes for $f=2.3$ shown in Fig.\ \ref{fig-avalanche-distr}(d) can be fitted using $\sigma=1$ with $a=0.0236\pm 0.0001$, $b=0.681\pm 0.002$, $d=87997\pm 1287$, and $a=0.0416\pm 0.0001$, $b=0.584\pm 0.002$, $d=87990\pm 912$, respectively, for the low and high driving rate, in the absence of site defects. When the site defects are present, corresponding to the curve with a  reduced cut-off size, we find $\sigma =1$, $a=0.0517\pm 0.0001$, $b=0.336\pm 0.002$, $d=106.01\pm 0.89$.
However, for the strong random-field disorder, Fig.\ \ref{fig-avalanche-distr}(c), a stretching exponent $\sigma >1$ is needed. Specifically, in the absence of site defects, we find $\sigma =1.68\pm 0.07$, $b=0.393\pm 0.005$, $d=74.3\pm 0.3$, and $\sigma =1.47\pm 0.02$, $b=0.319\pm 0.002$, $d=361\pm 3$ for low and high driving rate, respectively. In the presence of site defects and low driving rate, we find $\sigma =2.2\pm 0.3$, $b=0.41\pm 0.01$, $d=39.9\pm 1.4$. The parameter $a=0.11$ was kept fixed in all cases.

\begin{figure}[!htb]
\begin{tabular}{cc} 
\resizebox{18pc}{!}{\includegraphics{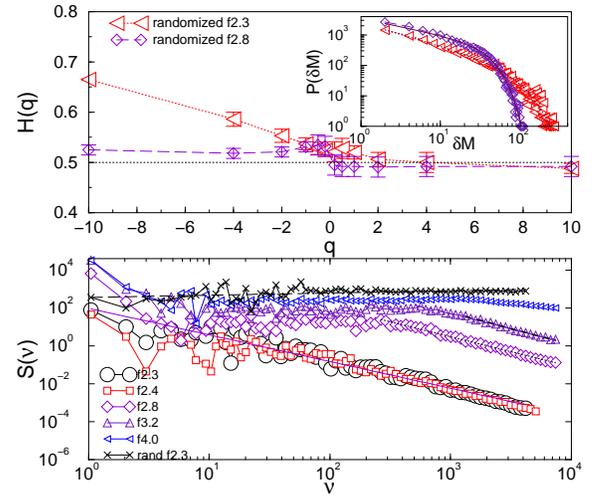}}\\
\end{tabular}
\caption{\small  (upper panel Generalized Hurst exponent $H(q)$ against  $q$ for two randomized Barkhausen noise signals in weak disorder, $f=2.3$, and strong disorder regime, $f=2.8$. Inset: The distributions of the individual jumps in these two cases. (lower panel) Power spectrum of Barkhausen time series for different disorder $f=2.3, 2.4, 2.8, 3.2, 4.0$, bottom to top,  and fixed driving rate $r=0.002$. The top line ($\times$) represents the spectrum of the randomized signal for $f=2.3$.}
\label{fig-PS}
\end{figure}

Apart from the observed scale invariance of the avalanches, the origin of the multifractality is often found in the underlying temporal correlations of the signal. In this respect, we determine the power spectrum of the Barhkausen time series for different pinning strengths, as shown in the lower panel of Fig.\ \ref{fig-PS}. The long-range correlations are manifested in the power-law dependence of the spectrum  $S(\nu )\sim \nu ^{-\phi}$ in a range of frequencies $\nu$. In particular, for low disorder we find $\phi =1.41\pm 0.04$ in a wide range of frequencies. On the other hand,  the exponent increased to $\phi=1.58\pm 0.05$ in the strong pinning regime and the range is gradually reduced towards high frequencies; straight lines indicate the fitted range. For comparison, the power spectrum of a randomized time series for $f=2.3$ is shown, top line, with the vanishing slope within the numerical error bars. The generalized exponent $H(q)$ corresponding to the randomized signal for $f=2.3$ and $f=2.8$ is shown in top panel of Fig.\ \ref{fig-PS}. The variation $\delta H$ is larger for the low disorder signal, $f=2.3$, suggesting that the variation in the data points obeys a non-exponential distribution. Indeed, the distributions of the data points, which are shown in the inset, confirm a broader range and a stretching of the cut-off in the case of low disorder signal. Stretched exponential distributions fit the data.

\section{Conclusions\label{sec-conclusions}}
We have studied the multifractal nature of the collective fluctuations in the magnetisation bursts (Barkhausen noise signals) originating from domain walls motion and pinning in the random-field Ising model on the hysteresis loop. First, the standard deviation of the fluctuation function shows an apparent dependence of the disorder or pinning strength. The resulting values of the Hurst exponent indicate that different classes of dynamical phenomena take place in the temporal fluctuations of the magnetisation burst $\{\delta M(k)\}$ at weak and strong disorder. In particular, the features of the fractional Brownian motion are recognised in the case of weak disorder, permitting movement of the individual domain walls. While properties of the fractional Gaussian noise characterise the strong disorder limit, where dense domain walls are present.  Further, the detrended multifractal analysis of the Barkhausen noise reveals the signal's complexity to a full extent. 
The generalised Hurst exponent $H(q)$ as a function of the parameter $q$ and the sizeable width of the spectrum $\delta H(q)>0$ appropriately quantify the observed multifractality. These measures indicate that the Barkhausen noise has different fractal features in various scales of fluctuations. Compared with the well studied hysteresis loop  criticality at the disorder-induced critical point with the exponents resembling the equilibrium critical phenomena, the multifractality of the Barkhausen noise occurs in the \textit{entire range of disorder} and a varied driving rate. Moreover, the observed multifractal features of the magnetisation fluctuations share a high similarity with complex signals in other physical, biological and social systems driven out of equilibrium. 

The sensitivity of the multifractal properties of the Barkhausen noise to the variations in the strength of disorder and driving rate can be directly related to the underlying collective  dynamics of the domain walls. Specifically, the increase of the exponents $H(q)$ for $q>0$ characterises the large fluctuations in the middle of the hysteresis loop in the case of weak disorder, where the motion and depinning of individual domain walls dominate the dynamics. In contrast, a decreasing function $H(q)$ is achieved in the strong disorder regime with many interacting domain walls. Generating new domain walls by structural defects on top of the weak random-field disorder confirm this picture. The presence of these other domain walls results with a decreasing $H(q)$ curve, as well as the reduced avalanches, as typically found in the strong disorder limit. 
The increased driving rates are manifested at most on the small-scale fluctuations by elevated values of the scaling exponents $H(q<0)$. Given the relation between the singularity exponent $\alpha $ and $H(q)$, the increased $H(q<0)$ values indicate a larger regularity of the smooth sections of the signal. On the other hand, a reduction of $H(q>0)$ appears (enhanced roughness) at large-scale fluctuations in comparison with the slow driving case. Although the broadening of the spectrum $\delta H$ due to fast driving is larger in the strong disorder,  the similar trend applies to the weak disorder case. 

In analogy with complex signals in many other driven systems, the Barkhausen  noise  possess strong signatures of the scale invariance in the avalanches and temporal correlations, non-Gaussian avalanche returns as well as a broad  distribution of the elementary pulses. These statistical features also depend on the strength of disorder and driving rates. However, the signal's multifractal properties appear to have a higher sensitivity to changed dynamical regimes, in particular with the combined variations of pinning and driving rates, as usually occur in the experiments.
Therefore, the multifractality of Barkhausen noise can be conveniently used for characterising the dynamical state of the domain walls in driven disordered systems on the hysteresis loop. These findings are relevant to the controlled motion of individual walls in recently investigated memory devices. Furthermore, here revealed multifractal structure of the Barkhausen noise highlights the importance of different scales in theoretical studies \cite{avalanche-evolution2013} of  the mechanisms of avalanche relaxation in a broader context of complex systems.

\acknowledgments
This work was supported by the program P1-0044 from the Research Agency of the Republic of Slovenia.


\end{document}